# An expanded X-ray beam facility (BEaTriX) to test the modular elements of the ATHENA optics

D. Spiga[1§], C. Pelliciari[1], E. Bonnini[2], E. Buffagni[2], C. Ferrari[2], G. Pareschi[1], G. Tagliaferri[1]

[1]INAF / Brera Astronomical Observatory, Via Bianchi 46, 23807 Merate  (Italy)
[2]CNR-IMEM, Parco Area delle Scienze 37/A – 43124 Parma - Italy

## ABSTRACT

Future large X-ray observatories like ATHENA will be equipped with very large optics, obtained by assembling modular optical elements, named X-ray Optical Units (XOU) based on the technology of either Silicon Pore Optics or Slumped Glass Optics. In both cases, the final quality of the modular optic (a 5 arcsec HEW requirement for ATHENA) is determined by the accuracy alignment of the XOUs within the assembly, but also by the angular resolution of the individual XOU. This is affected by the mirror shape accuracy, its surface roughness, and the mutual alignment of the mirrors within the XOU itself. Because of the large number of XOUs to be produced, quality tests need to be routinely done to select the most performing stacked blocks, to be integrated into the final optic. In addition to the usual metrology based on profile and roughness measurements, a direct measurement with a broad, parallel, collimated and uniform X-ray beam would be the most reliable test, without the need of a focal spot reconstruction as usually done in synchrotron light. To this end, we designed the BEaTriX (Beam Expander Testing X-ray facility) to be realized at INAF-OAB, devoted to the functional tests of the XOUs. A grazing incidence parabolic mirror and an asymmetrically cut crystal will produce a parallel X-ray beam broad enough to illuminate the entire aperture of the focusing elements. An X-ray camera at the focal distance from the mirrors will directly record the image. The selection of different crystals will enable to test the XOUs in the 1 - 5 keV range, included in the X-ray energy band of ATHENA (0.2-12 keV). In this paper we discuss a possible BEaTriX facility implementation. We also show a preliminary performance simulation of the optical system.

**Keywords:** ATHENA, X-ray telescopes, segmented X-ray optics, X-ray tests, expanded beam

## 1. INTRODUCTION

X-ray observatories of the future will be characterized by large apertures, on the order of 1 m radius and focal lengths of tens of meters. Among these ATHENA, to date selected[1] for the L2 slot in ESA's Cosmic Vision 2015–25 with a launch foreseen in 2028, is expected to address crucial topics in high-energy Astrophysics and Cosmology[2]. With a 3 m diameter optical module[3], a 12 m focal length, and an effective area of 2 m$^2$ at 1 keV, ATHENA will be the largest X-ray observatory ever built. In order to get high sensitivity and avoid source confusion, the angular resolution required for ATHENA is 5 arcsec HEW (Half Energy Width). With such large diameters it is not realistic to manufacture monolithic X-ray mirrors; the ATHENA optical module will therefore consist of nearly 1000 building blocks (named XOUs, X-ray Optical Units), each of them made by stacking some tens of the parabolic/hyperbolic lightweight segments with high focusing performances, kept properly spaced, and carefully aligned to reconstruct the widespread Wolter-I configuration[4]. These modular elements are to be carefully aligned into a supporting structure in order to attain the required 5 arcsec HEW. To this end, accounting for unavoidable alignment errors, the individual XOUs *must have resolutions much better than this limit*.

High angular resolution, lightweight modular optics are being developed at ESA/ESTEC and Cosine[5] since more than a decade, following the SPO (Silicon Pore Optics) technology[6], currently the baseline for the ATHENA optics. An alternative technology is based on SGO (Slumped Glass Optics), which have been successfully used to manufacture the optics of the hard X-ray telescope NuSTAR[7] and whose development is still ongoing at INAF/OAB[8] and MPE[9] (Garching, Germany). However, regardless of the adopted technique for such large optics, a mass production in an industrial facility is required. Because of unavoidable fabrication errors, quality controls have to be routinely performed at the industrial production site in order to discard the XOUs with insufficient angular resolution. However, owing to the dense stacking of the mirrors, it can be difficult to access the optical surfaces with the standard metrology tools. A direct characterization in X-rays, in contrast, provides direct information on the optical quality expected at the X-ray wavelength of operation. In addition, it offers the opportunity to carefully align the parabolic/hyperbolic segments of each XOU into the optimal configuration. X-ray tests of the SPOs are currently performed at the PTB laboratory of the BESSY synchrotron facility in pencil beam configurations[10] but they require a PSF reconstruction from each pore's exposure. In this way, performing functional tests of *all* the XOUs produced might be not so easy.

---

[§] contact author: Daniele Spiga, phone +39-039-5971027, email: daniele.spiga@brera.inaf.it

An efficient solution can be a full illumination test using a broad, low-divergent X-ray beam like the one in use at PANTER (MPE), now upgraded[11] to measure the Point Spread Function (PSF) optics with focal lengths up to 20 m. In fact, the extended PANTER facility was successfully employed in the last years to directly measure the angular resolutions of SPO and SGO prototypes[12] with a 20 m focal length. Nevertheless, it is convenient to use a large facility like PANTER to test large XOU assemblies (e.g. petals) of ATHENA or other telescopes, but not to routinely perform functional tests of more than 1000 XOUs, not to mention the less performing XOUs that will not pass the quality check.

The problem of producing a parallel, collimated X-ray beam to illuminate the entire aperture of the XOUs that will be produced for ATHENA is that the X-ray source needs to be located at very large distance. In other words, the source divergence must be very low to avoid the finite distance effects that affect the on-ground characterization of a Wolter-I optic[4]. For instance, if $S$ is the available source-detector distance, the nominal focal distance $f$ is shifted[13] to

$$f' = \frac{S}{2}\left(1 - \sqrt{1 - \frac{4f}{S}}\right).$$

This implies that, if $S < 4f$, then the angular resolution cannot be directly measured in focus. Therefore, in the case of ATHENA the source has to be set at $S \gg 50$ m to enable an affordable measurement. This requirement entails a very large volume in vacuum and a large facility size that can be hardly accommodated in an industrial production chain. For example, at PANTER the low divergence is made possible by $S \approx 130$ m (with a further collimation that is being implemented via a diffracting zone plate[14]).

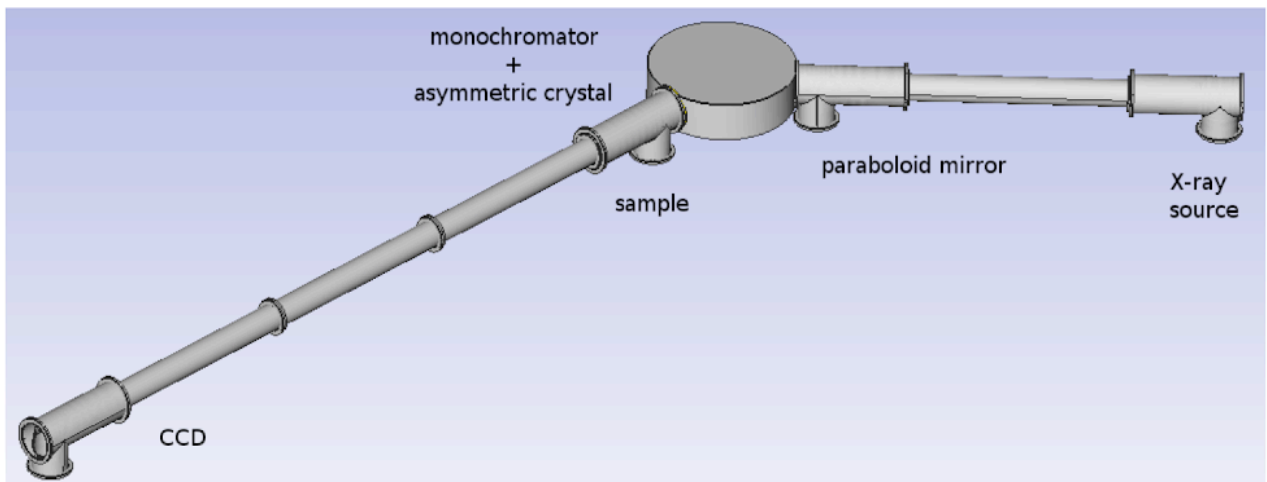

Fig. 1: outer view of the BEaTriX facility project.

In this paper we present a project of a small X-ray facility (BEaTriX, *Beam Expander Testing X-ray facility*) to perform the functional tests of the ATHENA XOUs. The design foresees the collimation and the expansion of an X-ray beam produced by a source located at a short distance (~ 5 m) from the module under test. The wavefront expansion and collimation will be achieved via diffraction onto an asymmetric cut crystal[15], a concept already implemented successfully at the Daresbury synchrotron[16]. However, that setup was expanding the beam in only one direction. For a beam expansion in 2 dimensions, we propose an extended setup including a parabolic mirror, a monochromator and an asymmetric cut crystal. The short range of the rays does not require operating in high vacuum, which makes the facility operational in a short time, and the characterization of a single module a matter of a few hours, therefore suitable to sustain a high production rate. In the next section we describe briefly the BEaTriX project and give an overview of the optical elements properties. In Sect. 3 we deal with some preliminary performances simulations, aiming at demonstrating the capabilities of the apparatus to be built at INAF/OAB, and in a subsequent time to be replicated at the production site. We draft some conclusions in Sect. 4.

## 2. OVERVIEW OF THE BEATRIX PROJECT

The BEaTriX facility project is conceived to perform in a compact space *the functional tests of the X-ray optical units of a composite optical module for a large X-ray telescope like ATHENA*. The apparatus will produce a broad (200 mm × 60 mm), uniform, and parallel (< 2 arcsec) beam of X-rays. These requirements are needed to illuminate the full aperture of the largest XOUs of ATHENA and to measure their HEW to an accuracy of ~2 arcsec or better.

Viewed from outside, (Fig. 1) BEaTriX will comprise two arms: the "short" arm, approx. 5 m long, will contain and keep under low vacuum the expanding/collimating elements for the X-rays created by a commercial X-ray source. The long arm will enable the propagation of focused X-rays from the XOU toward a CCD detector, and its length will equal the XOU focal length (as of today, 12 m). Both arms will be rotated to fit the nominal incidence angle of the XOU sample to be tested, which will be housed in the cylindrical box at the intersection of the two arms.

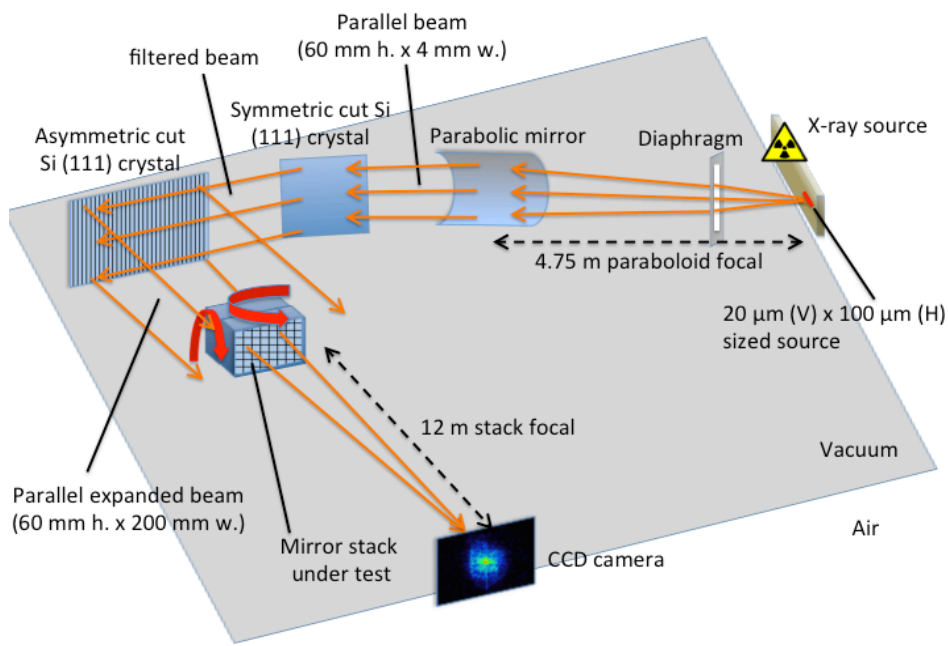

Fig. 2: a scheme of the BEaTriX facility project. The entire system, but the source and the camera, is kept in low vacuum (0.1 mbar).

The optical layout of BEaTriX is shown in Fig. 2. The X-ray source can be a commercial X-ray tube, providing the usual bremsstrahlung continuum and the fluorescence lines of the anode material. We have selected two possible X-ray energies corresponding to different BEaTriX settings: at 1.49 keV (the Al-Kα line) or 4.51 keV (the Ti-Kα line), both in the energy band of ATHENA (0.2-12 keV). In order to ensure the correct beam collimation, the electron spot on the anode, seen on a plane normal to the line of sight, must not exceed 20 μm in the vertical direction and 100 μm in the horizontal one. The small vertical size is needed to keep the vertical divergence below 1 arcsec at a quite short distance to the XOU (~5 m). In the horizontal direction, the source size is less critical because the subsequent asymmetric diffraction (Fig. 5) collimates the diffracted beam to within a couple of arcseconds (Fig. 6). The asymmetric diffraction efficiency is quite low, but this is not a problem in general because even commercial X-ray sources are characterized by a copious flux[**] (as large as to ~5×$10^6$ counts/sec/$deg^2$).

After a preliminary slit to limit the disturbance of stray light, X-rays propagate inside the short arm of the facility, in low vacuum (< 0.1 mbar). In Fig. 3 we report the calculated transmittance of the total range of rays from the source to the CCD: even for a residual pressure of 0.1 mbar, only 15% of the flux at 1.49 keV is absorbed, while the attenuation at 4.51 keV is almost negligible.

The first collimation stage is a grazing incidence mirror shaped as a paraboloidal sector (Fig. 4, left), with the X-ray source in its focus, which parallelizes, collimates, and expands the X-ray beam in the vertical direction to a 60 mm size. The sole mirror also expands the beam in the horizontal direction, but only to a 4 mm size (Fig. 4, right).

---
[**] http://www.amptek.com/products/

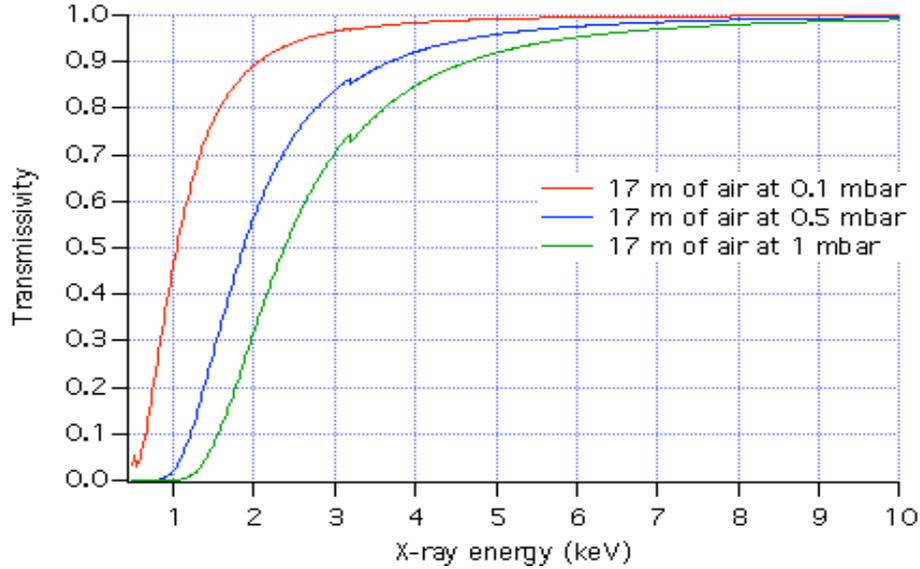

Fig. 3: transparence of a 17 m column of low vacuum. For a residual pressure of 0.1 mbar, the attenuation is low also at 1.49 keV.

To ensure a proper performance of BEaTriX, the surface quality of the collimating mirror (HEW < 1 arcsec) is crucial. Figure errors or a non-negligible surface roughness scatter the rays, mostly in the incidence plane. In addition to degrading the horizontal collimation, this would also affect the incidence angle uniformity on the diffracting crystals, making the final beam non-uniform. A suitable material to manufacture the mirror can be Zerodur[TM], owing to its extremely low thermal expansion, and because it can be polished to an excellent level. To reduce the costs, the collimating mirror can be figured from a purchased conical segment, using one of the two Ion Beam Figuring machines operated at INAF/OAB[17],[18]. Also the mirror polishing can be achieved with a lapping machine recently developed in our labs. Eventually, a 30 nm Platinum coating plus a 5 nm amorphous Carbon over-coating[19] endows the mirror with a reflectivity of 89% at 1.49 keV and of 41% at 4.51 keV.

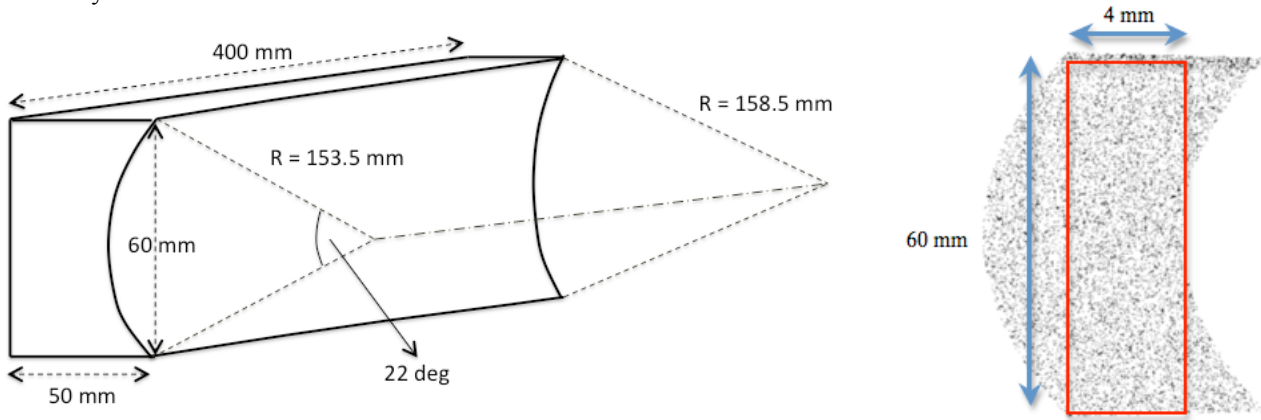

Fig. 4: (left) scheme of the parabolic mirror used to expand the beam in the vertical direction. The focal length is 4.75 m and the grazing incidence angle is 0.93 deg. The beam emerges from the mirror shaped alike the figure on the right side (the horizontal scale is stretched). The red rectangle is the part that, after the monochromation, impinges on the asymmetric crystal.

Just after the exit of the collimating mirror, the collimated beam is diffracted in sequence by a couple of crystals. For the 4.51 keV ($\lambda$ = 2.7 Å) setup, Silicon crystals can be adopted using the (111) diffraction, characterized by a d-spacing of 3.13 Å. The first crystal is symmetrically cut, i.e. parallel to the Bragg planes; therefore, the diffracted beam is flipped horizontally, but remains unchanged in shape and size. The symmetrically cut crystal, set at the correct Bragg angle of $\varphi_B$ = 26.35 deg from the surface, filters the $E$ = 4.51 keV fluorescence line from the continuum, thereby improving the final collimation in the subsequent diffraction. A computed rocking curve of the symmetric crystal is shown in Fig. 5 (red line). The rocking curve FWHM is $\Delta\varphi_B$ = 15 arcsec, which corresponds to a spectral acceptance $\Delta E = E\,\Delta\varphi_B\,/\,\tan\varphi_B$ = 0.5 eV, close to the spectral width of a typical fluorescence line.

Once monochromated, the asymmetrically cut crystal finally expands the beam horizontally to a 200 mm size (Fig. 6). Unlike the previous crystal, the outer surface forms an angle $\alpha = 25.5$ deg with the (111) Bragg planes. Hence, if the beam impinges at the Bragg angle, it forms an angle $\varphi_{in} = \varphi_B - \alpha = 0.85$ deg from the surface and the 4 mm wide beam is uniformly spread over the crystal length of 260 mm. The reflection occurs with respect to the Bragg planes, so that the beam exits at $\varphi_{out} = \varphi_B + \alpha = 51.85$ deg angle from the surface. Owing to this asymmetry (Fig. 6) the beam width is amplified by a 53-fold factor. This beam area gain is compensated, in accordance to Liouville's theorem, by a rocking curve 53 times narrower in reflection angles than in the incidence angles, and also much narrower than the rocking curve in the symmetric diffraction (Fig. 5). This shrinkage of the rocking curve width definitely improves the beam collimation to a value suitable to characterize the ATHENA XOUs, even if the horizontal size of the source is not very small. On the other hand, the range of incidence angles that can be accepted by the asymmetric crystal is 100 arcsec wide (FWHM), hence the horizontal collimation before the asymmetric diffraction must not necessarily be very high in order to maintain high diffraction efficiency.

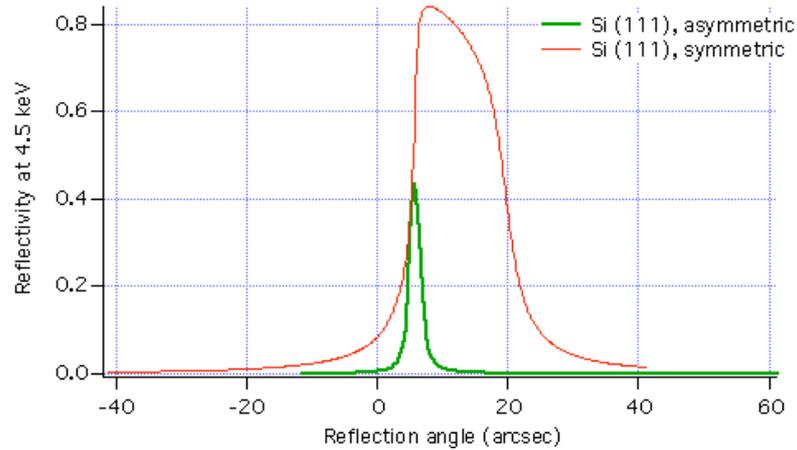

Fig. 5: rocking curves for a Silicon crystal at 4.51 keV in symmetric and asymmetric diffraction (both computed with the SHADOW code[††]). The symmetric diffraction is not effective in collimation capability because the rocking curve is some tens arcsec wide. In contrast, the asymmetric diffraction returns a beam collimated within a couple of arcseconds of FWHM.

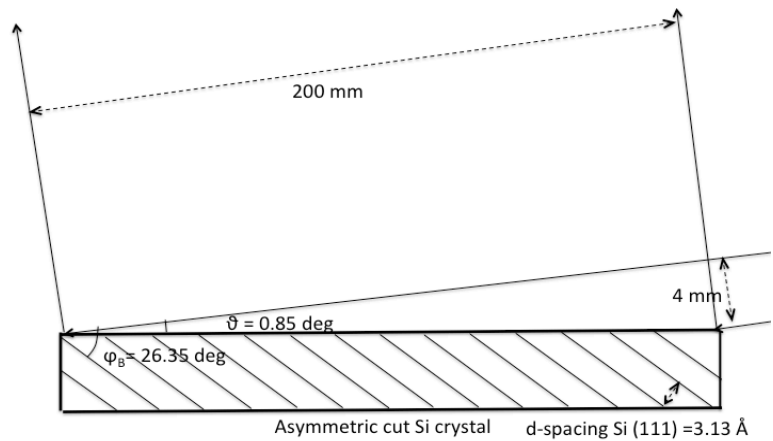

$$M = \frac{\sin(\varphi_B + \alpha)}{\sin(\varphi_B - \alpha)} = 53 \approx \frac{200 mm}{4 mm} \qquad \varphi_B \text{ (Bragg Angle)} > \text{asymmetry angle } \alpha$$

Fig. 6: scheme of the asymmetric reflection on the Silicon crystal used to expand horizontally the X-ray beam. The crystal has a 260 mm length. Since the rays strike on the outer surface crystal at the shallow angle $\varphi_B - \alpha$ but are reflected at $\varphi_B + \alpha$, the diffracted beam is uniformly expanded by a factor of ~50 in the horizontal direction, i.e., to a 200 mm width.

---

[††] http://www.esrf.eu/computing/scientific/raytracing/

Silicon crystals can be purchased from different industrial providers to a very high level of lattice perfection (better than 0.1 arcsec). The symmetric/asymmetric cut according to the correct angles with respect to the selected crystalline planes, and the subsequent surface polishing, will be done at IMEM-CNR laboratories.

For the setup at 1.49 keV ($\lambda$ = 8.3 Å), Si (111) planes cannot be used, because the Bragg law $\sin\varphi = \lambda/2d$ cannot be fulfilled. We have evaluated the possibility to replace the crystals with planar reflection gratings, but in order to achieve in the asymmetric configuration the needed beam expansion ($M \approx 50$, Fig. 6) the entrance angle, $\varphi_{in}$, and the exit angle, $\varphi_{out}$, need to be very different; hence, the line step at the first diffraction order would have to be

$$d = \frac{\lambda}{\cos\varphi_{in} - \cos\varphi_{out}} = 2 \div 3 \text{ nm}$$

over a length of 260 mm. This is beyond the capabilities of most grating rulers, while the most advanced EUV photolithographic processes can reach performances down to 30 nm pitches, and at not so affordable costs. Higher orders of diffraction (say, the 10$^{th}$) would allow us to use line steps of this order of magnitude, but the grating efficiency would be extremely low (< 0.1%), much lower than a crystal (Fig. 5). Hence, the ideal solution for the 1.49 keV setup would be the adoption of a crystal with a larger elementary cell, like Lithium Niobate (NbLiO$_3$, c = 13.9 Å), also commercially available with extremely low mosaicity. Also in this case, the crystals can be purchased in chunks of a proper size, and then be cut with the correct orientation and polished at IMEM-CNR.

Finally, the collimated, uniform, parallel, and 200 mm × 60 mm beam illuminates the sample that can be aligned via precise stepper motors. After the reflection, the beam propagates in the 12 m tube and is focused onto a CCD camera, enabling thereby the direct characterization of the tested XOU PSF, from which the angular resolution HEW can be calculated immediately. Performing characterization of the same XOU at 1.49 keV and 4.51 keV provides some trend of angular resolution throughout the ATHENA energy band (0.2-12 keV) and allows us to disentangle the effects of the XOU roughness (increasing with the X-ray energy) from the XOU profile error (independent of the X-ray energy). Possible extensions to higher energies might be also envisaged.

## 3. SYSTEM SIMULATION

We have simulated the collimation stage of BEaTriX using a code written in IDL language. The code traces rays from the 4.51 source (100 µm h. × 20 µm v.) in random directions within the entrance pupil of the parabolic mirror (Fig. 7). The reflection off the parabolic mirror is computed according the geometric optics with a survival probability equal to the coating reflectivity. The surviving rays are traced to the symmetric crystal and then to the asymmetric one, reflected in both cases with likelihood depending on their incidence angle, according the reflectivities shown in Fig. 5.

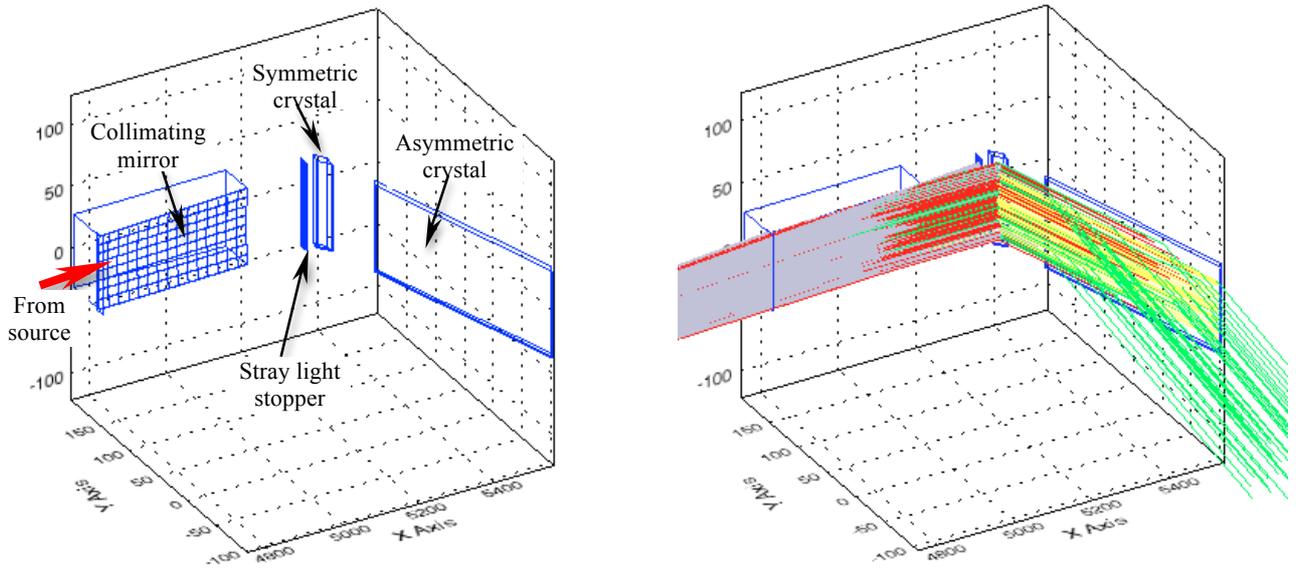

Fig. 7: (left) arrangement of the BEaTriX optical elements at 4.51 keV. (right) tracing X-ray throughout BEaTriX: in gray, the rays that are blocked. In red, the rays absorbed. Stray rays are depicted in yellow and rays that have reached the focal plane (Fig. 10) are green. The x-axis scale is compressed with respect to the other two.

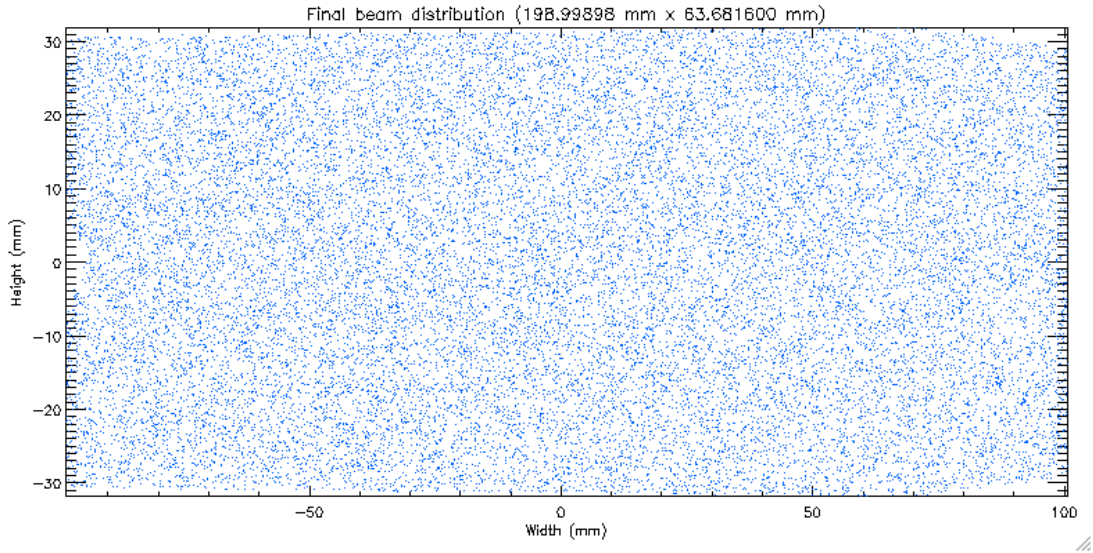

Fig. 8: simulated "flat field", i.e., final positions of the diffracted rays over plane at a 12 m distance (the focal length of ATHENA) away from the asymmetric crystal. Without a focusing module, the area is covered uniformly. 200000 rays were launched, only 4% have reached the target. The others were absorbed in either the mirror, or the crystals, or were blocked as stray light, or (very few of them) absorbed in the residual atmosphere. The 200 x 60 mm area is covered uniformly.

The surviving rays are then propagated out to the CCD camera: since the focusing XOU is not included in the simulation, the beam remains as wide as the projected size of the asymmetric crystal, so it has to be recorded on a screen of the same size. At every step of the computation, the absorption probability in the residual atmosphere is evaluated and the rays are propagated or stopped accordingly. The final result is that 4% of the rays traced have reached the CCD plane.

As a performance metric, we obtain an expanded beam of 200 mm × 60 mm size uniformly covered with collimated rays (Fig. 8). Accounting for the estimated intensity of available X-ray sources (Sect. 2), the achievable density is 30 to 80 counts/sec/cm$^2$ (depending on the exact characteristics of the source), sufficient to provide a fast PSF characterization of the XOU. The mentioned performance refers to flawless optical components, especially the collimating mirror that has to be very accurate in shape and roughness. Real components will exhibit defects that may degrade the uniformity and the collimation of the expanded beam. These defects will be included in the simulation to determine the fabrication tolerances of the components.

## 4. CONCLUSIONS AND FUTURE WORK

The BEaTriX facility will have the capability to perform *a direct, non-destructive characterization of single and stacked integrated mirrors as modular elements of future large area X-ray telescopes.* It will enable the selection of the stacks with the best angular resolution (regardless of the XOU manufacturing technique) in a fast and reliable way:
- A full-illumination characterization will be performed without suffering from the finiteness of the source distance.
- The small size of the tank and the low vacuum required will make the venting/sample change/pumping down a few hours matter. Hence, a large number of samples will be tested in a given time and a quicker feedback after the manufacturing will be provided.
- It will also enable the alignment of the parabolic/hyperbolic stacks under X-rays.
- The characterization will be *in situ* and *in real time*.

Finally, for mirrors with a different focal length, the facility can be adapted easily by adding or removing (and trimming if needed) segments of the 12 m long arm, between the optic and the camera. More simulations will be performed to simulate the effects of the imperfections and the misalignments of the optical elements in order to set manufacturing and mounting tolerances for BEaTriX, before starting the production of the components.

## ACKNOWLEDGMENTS

We thank Manuel Sanchez Del Rio (ESRF, Grenoble, France), Finn E. Christensen (DTU, Danemark), and Marco Barbera (Università di Palermo, Italy) for useful discussions.